%% file: main.tex
\documentclass[twocolumn]{svjour3}

\input{preamble}

\smartqed  
\journalname{Software and System Modeling}

\begin{document}

\placetextbox{0.5}{0.99}{\large\colorbox{gray!10}{\textcolor{WildStrawberry}{\textbf{Author pre-print.}}}}%

\placetextbox{0.5}{0.97}{\large\colorbox{gray!10}{\textcolor{WildStrawberry}{\textbf{The final publication is available at: \url{https://doi.org/10.1007/s10270-022-01054-5}.}}}}%

\placetextbox{0.5}{0.05}{\colorbox{gray!10}{\textcolor{WildStrawberry}{\textbf{Author pre-print. The final publication is available at: \url{https://doi.org/10.1007/s10270-022-01054-5}.}}}}%

\title{Real-time Collaborative Multi-Level Modeling by Conflict-Free Replicated Data Types}

\author{
    Istvan David \and Eugene Syriani
}

\authorrunning{I. David and E. Syriani}

\institute{DIRO -- Université de Montréal, Canada, \email{istvan.david@umontreal.ca, syriani@iro.umontreal.ca}}

\date{Received: date / Accepted: date}

\maketitle

\input{sections/abstract}
\input{sections/intro}
\input{sections/casestudy}
\input{sections/background}
\input{sections/framework}
\input{sections/elaboration}
\input{sections/discussion}
\input{sections/conclusions}
\input{sections/acknowledgments}

\bibliographystyle{spbasic}
\bibliography{bib/mdse,bib/crdt,bib/other}


\end{document}

%% file: preamble.tex
\input{preamble/packages}
\input{preamble/commands}
\input{preamble/settings}
\input{preamble/listing}

%% file: preamble/packages.tex
\usepackage[utf8]{inputenc}
\usepackage[T1]{fontenc}
\usepackage{ifthen}
\usepackage{hyperref}

\usepackage[dvipsnames]{xcolor}
\usepackage{soul}
\usepackage[normalem]{ulem}

\usepackage{epsf,picinpar}
\usepackage{varioref}
\usepackage{fdsymbol}

\usepackage{pifont}
\usepackage{fancyhdr}

\usepackage[numbers]{natbib}

\usepackage{subcaption}

\usepackage{harveyballs}
\usepackage{graphicx}
\usepackage{xspace}
\usepackage[strings]{underscore}
\usepackage{enumitem}
\usepackage{booktabs}
\usepackage{rotating}
\usepackage{multirow}
\usepackage[pscoord]{eso-pic}

%% file: preamble/commands.tex
\newcommand{\secref}[1]{Section~\ref{#1}}

\newcommand{\figref}[1]{Fig.~\ref{#1}}
\newcommand{\tabref}[1]{Table~\ref{#1}}

\newcommand{\lstref}[1]{Listing~\ref{#1}}

\newcommand{\todo}[1]{
    \ifthenelse{\equal{#1}{}}{\textcolor{red}{TODO}}{\textcolor{red}{TODO:~{#1}}}
}

\newcommand{\rem}[1]{%
    \ifthenelse{\boolean{showannotations}}%
    {\textcolor{red}{\st{#1}}}%
    {}%
}

\newcommand\add[1]{%
    \ifthenelse{\boolean{showannotations}}%
    {\textcolor{\newtextcolor}{{#1}}}%
    {#1}%
}

\newcommand\rep[2]{%
    \ifthenelse{\boolean{showannotations}}%
    {\rem{#1}~\add{#2}}%
    {#2}%
}

\newcommand{\lowkey}{{\fontfamily{lmss}\selectfont lowkey}}

\newcommand{\lwwregister}{\texttt{LWWRegister}}
\newcommand{\lwwset}{\texttt{LWWSet}}
\newcommand{\lwwmap}{\texttt{LWWMap}}
\newcommand{\lwwgraph}{\texttt{LWWGraph}}
\newcommand{\lwwvertex}{\texttt{LWWVertex}}
\newcommand{\lwwedge}{\texttt{LWWEdge}}

\newcommand{\ie}{i.e.,\xspace}
\newcommand{\eg}{e.g.,\xspace}

\newcommand{\operation}[1]{\phantom{}\\\noindent\textbf{#1}}

\newcommand{\placetextbox}[3]{
  \setbox0=\hbox{#3}
  \AddToShipoutPictureFG*{
    \put(\LenToUnit{#1\paperwidth},\LenToUnit{#2\paperheight}){\vtop{{\null}\makebox[0pt][c]{#3}}}%
  }%
}%

%% file: preamble/settings.tex
\newboolean{showannotations}
\setboolean{showannotations}{true} 

\newboolean{shownames}
\setboolean{shownames}{true}

\newcommand{\newtextcolor}{blue}

\newcommand*\rot{\rotatebox{0}}

%% file: preamble/listing.tex
\usepackage{listings}
\usepackage{color}

\definecolor{darkgreen}{rgb}{0,0.6,0}
\definecolor{gray}{rgb}{0.5,0.5,0.5}
\definecolor{mauve}{rgb}{0.58,0,0.82}
\definecolor{gray}{rgb}{0.4,0.4,0.4}
\definecolor{darkblue}{rgb}{0.0,0.0,0.6}
\definecolor{lightblue}{rgb}{0.0,0.0,0.9}
\definecolor{cyan}{rgb}{0.0,0.6,0.6}
\definecolor{darkred}{rgb}{0.6,0.0,0.0}
\definecolor{armygreen}{rgb}{0.29, 0.33, 0.13}
\definecolor{antiquefuchsia}{rgb}{0.57, 0.36, 0.51}
\definecolor{coolblack}{rgb}{0.0, 0.18, 0.39}

\lstdefinelanguage{XML}
{
    morestring=[s][\color{antiquefuchsia}]{"}{"},
    morestring=[s][\color{black}]{>}{<},
    morecomment=[s]{<?}{?>},
    morecomment=[s][\color{darkgreen}]{<!--}{-->},
    stringstyle=\color{black},
    identifierstyle=\color{darkred},
    keywordstyle=\color{darkred},
    morekeywords={xmlns,xsi,noNamespaceSchemaLocation,type,id,x,y,source,target,version,tool,transRef,roleRef,objective,eventually},
    emph=[1]{name,implicit,grammarName,grammarType,library},
    emphstyle=[1]{\color{armygreen}}
}

\lstdefinelanguage{Xtext}
{
    morestring=[s][\color{darkred}]{'}{'},
    identifierstyle=\color{black},
    morekeywords={def,FOR,ENDFOR,IF,ENDIF},
    keywordstyle=\color{lightblue},
    emph=[1]{Function,RFID125,BranchRFID125,ReadCodeRFID125,WriteCodesRFID125,Concept,Attribute,SubType,Reference,Containment,Textual,Graphical,AttributeType},
    emphstyle=[1]{\color{armygreen}},
    emph=[2]{pin1,pin2,pin,codes,attributes,relation,super,ref_from,container,attrType,representation,conceptName},
    emphstyle=[2]{\color{antiquefuchsia}},
    emph=[3]{STRING,ID},
    emphstyle=[3]{\color{coolblack}},
    extendedchars=true
}

\lstdefinelanguage{Command}
{
    comment=[l]{--},
    morekeywords={READ,OBJECTS,CREATE,LINK,UPDATE,DELETE,TO},
    keywordstyle=\color{darkblue}\bfseries,
    morestring=[s][\color{darkred}]{\{}{\}},
    extendedchars=true
}

%% file: sections/abstract.tex
\begin{abstract}
The need for real-time collaborative solutions in model-driven engineering has been increasing over the past years.
Conflict-free replicated data types (CRDT) provide scalable and robust replication mechanisms that align well with the requirements of real-time collaborative environments.
In this paper, we propose a real-time collaborative multi-level modeling framework to support advanced modeling scenarios, built on a collection of custom CRDT, specifically tailored for the needs of modeling environments. 
We demonstrate the benefits of the framework through an illustrative modeling case and compare it with other state-of-the-art modeling frameworks.

\end{abstract}

\keywords{Collaborative modeling, Real-time collaboration, Multi-level modeling, Conflict-free replicated data types, Model-driven engineering}

%% file: sections/intro.tex
\section{Introduction}\label{sec:introduction}

Collaborative Model-Driven Software Engineering (MDSE) \citep{muccini2018collaborative} aims to establish a sound interplay between physically distanced stakeholders by combining the techniques of collaborative software engineering \citep{whitehead2007collaboration} and model-driven techniques \citep{schmidt2006model}.
Recent systematic studies \citep{david2021collaborative,franzago2018collaborative} report a substantial shift towards real-time collaboration in MDSE. While real-time collaborative MDSE opens up many opportunities, it also gives rise to unique challenges, primarily: ensuring appropriate convergence of distributed data, while still guaranteeing timely execution \citep{sun1998achieving}.

Optimistic replication has been suggested by \citet{saito2005optimistic} as a possible treatment, in which the local replicas of stakeholders are allowed to diverge temporarily. This divergence is admissible due to eventual consistency mechanisms \citep{vogels2009eventually} ensuring that each remote change will be observed by every stakeholder eventually and replicas converge. Strong eventual consistency (SEC) \citep{preguicca2009commutative} augments eventual consistency with a safety guarantee: two replicas that have received the same set of change updates will be in the same state, regardless of the order of updates. Although SEC aligns well with the requirements of real-time collaborative MDSE settings, it is not trivial to implement correctly \citep{deporre2019putting}.

Conflict-free Replicated Data Types (CRDT) have been suggested by Shapiro et al.~\cite{shapiro2011conflict} as a scalable implementation of SEC.
CRDT have been traditionally geared to support linear data, such as text. Since traditional software engineering relies on textual artifacts to persist source code, mechanisms of real-time collaborative \textit{textual} editors can address the main challenges of real-time collaborative source code development.
However, MDSE relies on richer data types, such as multigraphs, that are also potentially disconnected. Thus, CRDT cannot efficiently accommodate \textit{models} as first-class citizens.
Existing solutions either (i) focus exclusively on textual modeling and reduce collaboration to textual primitives which are well-supported by current CRDT frameworks \citep{saini2021towards}; or (ii) work on models of limited complexity and do not support proper graph semantics at the data level \citep{derntl2015near}.
As a consequence, current CRDT-based techniques fall short of supporting intricate modeling scenarios.

Such an intricate modeling scenario is multi-level modeling (MLM)~\cite{atkinson2001essence}. MLM attempts to overcome the limitations of traditional modeling architectures, such as the OMG's Meta-Object Facility (MOF)\footnote{\url{https://www.omg.org/mof/}}~\cite{atkinson2002rearchitecting,atkinson2008reducing,gonzalezperez2008metamodelling}, stemming from their restricted, two-level meta-modeling approach. While two meta-layers might be enough for simple modeling cases, such architectures fail to support scenarios in which different kinds of conformance might be required, or in which language designers need the same level of control over indirect instances two or more meta-levels below~\cite{delara2010deep}. Misusing two-level modeling architectures to emulate multi-level modeling gives rise to undesirable modeling problems. Such cases have been reported by \citet{brasileiro2016applying} who point out that 85\% of wiki knowledge base classes allow for the inference of unsound conclusions.
To react to these issues, MLM approaches increase the flexibility of the meta-modeling architecture by allowing an arbitrary number of meta-levels~\cite{delara2010deep}.

Augmenting multi-level modeling with real-time collaboration capabilities enables teamwork between stakeholders acting (i) at different levels of abstraction, or (ii) at different levels of decision making. Typical examples include (i) changing a modeling language during operation to incorporate language elements required by the technical stakeholders \citep{izquierdo2016collaboro}; and (ii) restricting values of attributes at higher levels of decision making and enforcing these values through the notion of potency \citep{van2014multi}.
However, current collaboration frameworks are limited to two-level modeling architectures and no sound practices exist to engage in real-time collaborative multi-level modeling among distributed stakeholders~\cite{david2021collaborative,franzago2018collaborative}.

\paragraph{Contributions.}
The main contribution of our work is a novel real-time collaborative multi-level modeling framework, called \lowkey{}\footnote{\url{https://github.com/geodes-sms/lowkey}}.
The framework supports a wide range of modeling scenarios across an arbitrary number of modeling and linguistic meta-levels. We achieve this flexibility by separating the linguistic metamodel(s) from the physical metamodel. In this setting, domain-specific models always conform to their linguistic metamodels, and they both (models and metamodels) conform to the uniform physical metamodel provided by the framework. Real-time synchronization between collaborating stakeholders is achieved by persisting the physical metamodel in CRDT that ensure the consistency of domain-specific models in a domain-agnostic fashion. This makes \lowkey{} especially suitable for supporting approaches such as multi-view modeling \citep{reineke2014basic}. Furthermore, \lowkey{} subsumes traditional modeling frameworks, such as UML \citep{fowler2004uml} and EMF \citep{steinberg2008emf}, offering a promising integration potential with existing model editors.

\paragraph{Structure.}
The rest of this paper is structured as follows.
First, in \secref{sec:case}, we provide an example to illustrate the concepts throughout the paper.
In \secref{sec:background}, we give an overview of the background of our work, and review the related work.
In \secref{sec:framework}, we present \lowkey{},  our framework for real-time collaborative modeling.
In \secref{sec:elaboration}, we demonstrate the feasibility of the framework by applying it on the case study.
In \secref{sec:discussion}, we discuss the lessons learned and compare \lowkey{} with similar frameworks.
Finally, in \secref{sec:conclusions}, we draw the conclusions.

%% file: sections/casestudy.tex
\section{Illustrative case}\label{sec:case}

We rely on an illustrative case of developing a collaborative editor for modeling Mind maps \citep{Buzan2006}. The case highlights multiple structural facets of (meta)modeling, such as typing, inheritance, and various forms of well-formedness through the collaborative modeling of mind maps. We show how an editor built on top of \lowkey{} would enable this.

\begin{figure}
    \centering
    \includegraphics[width=\linewidth]{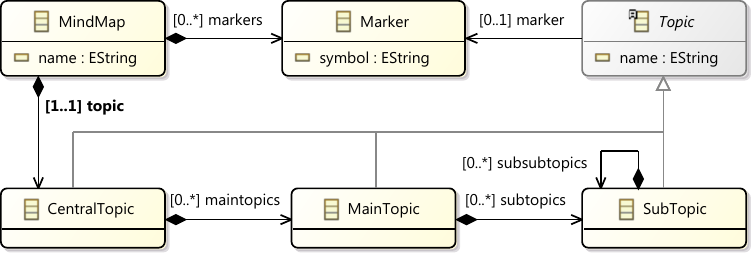}
    \caption{The metamodel of the Mind maps.}
    \label{fig:mindmapmm}
\end{figure}

The metamodel of the case is shown in~\figref{fig:mindmapmm}. \textit{MindMap} serves as the root element, containing the various types of \textit{Topic}s, and the \textit{Marker}s that can be associated with specific \textit{Topic}s. The mindmap is hierarchical: the single \textit{CentralTopic} further contains an arbitrary number of \textit{MainTopic}s, each containing an arbitrary number of \textit{SubTopics}. Finally, \textit{SubTopics} can contain \textit{SubTopics} to arbitrary depths.

Based on the metamodel, the language engineer must provide an editor that enables real-time collaborative modeling of a \textit{MindMap} instance. This entails the following operations: creating, editing, and deleting instances of metamodel concepts; creating, editing, and deleting links between elements. In addition, a mechanism for reading the state of the local model is required.

\paragraph{Scenarios.}
Here are some modeling scenarios that may occur as a result of applying these CRUD operations.

\begin{description}
    \item[\textbf{Cooperation.}] User A creates a \textit{MindMap} instance and User B sets its \textit{title}.

    \item[\textbf{Cooperation with of linguistic inconsistencies.}] \phantom{}\\ User A creates a \textit{CentralTopic} instance, but does not link it to the \textit{MindMap} instance. User B links the \textit{CentralTopic} instance to the \textit{MindMap} via a composition reference.

    \item[\textbf{Conflict management.}] User B removes the \textit{CentralTopic} instance; in parallel, however, User A creates a \textit{MainTopic} instance and links it to the \textit{CentralTopic} instance.

    \item[\textbf{Multi-level cooperation.}] Users A and B have created multiple \textit{Marker}s, and would like to categorize them (e.g., as textual and graphical). The Language Designer, who works one metalevel above Users A and B, sets the \textit{potency} of the \textit{Marker} type from 1 to 2; thus, allowing a templating mechanism for A and B.
    
    \item[\textbf{Multi-level integrity.}] Subsequently, the Language Designer decides to revoke this level of flexibility from the users and sets the potency of the \textit{Marker} class to 1, rendering the inheritance chain inconsistent.

\end{description}

\paragraph{Requirements.}
Based on the requirements for effective real-time collaboration defined by Sun et al.~\cite{sun1998achieving}, we formulate the following requirements for a solution in the context of this illustrative case. These requirements are evaluated in the feasibility study in \secref{sec:elaboration}.
\begin{enumerate}[label={R}{{\arabic*}}.,leftmargin=0.7cm]
    \item Convergence, \ie every stakeholder’s local data must exhibit the same state after updates have been applied.
    \item Timely execution, i.e., operations must propagate and the system must reconcile within a deadline to provide a smooth user experience on the client side.
    \item Causality preservation, i.e., updates must be ordered in the same causal way by each stakeholder.
    \item User intention preservation, i.e., the original user’s intention must be preserved.
    \item Multi-level collaboration, i.e., ability to collaborate and resolve inconsistencies across different meta-levels.
    \item Meta-model reuse, i.e., tool builders have to be able to use previously defined meta-models with only minimal adaption.
    \item Tolerance of linguistic inconsistencies, i.e., the ability to synchronize and persist models into the physical representation despite linguistic non-conformance of model elements.
\end{enumerate}

%% file: sections/background.tex
\section{Background}\label{sec:background}

We discuss the existing work related to collaborative modeling.

\subsection{Collaborative MDSE}\label{sec:background-collaborativemodeling}

Collaborative software engineering enables effective cooperation among stakeholders~\citep{whitehead2007collaboration}, often in distributed settings~\citep{herbsleb2007global}. Distributed teams introduce challenges to collaboration in terms of processes, project management, artifact sharing, and consistency~\citep{mistrik2010collaborative}.
These challenges are further exacerbated in the engineering of complex software-intensive systems that require collaboration between stakeholders of highly diverse expertise. MDSE~\citep{schmidt2006model} provides stakeholders with techniques for reasoning about the system at higher levels of abstraction than source code.
As the combination of collaborative software engineering and MDSE, collaborative MDSE exhibits the traits of both disciplines.
Collaborative MDSE has become a prominent feature of nowadays' software engineering practice~\citep{brambilla2017model}.
Version control for modeling artifacts
has been extensively employed to facilitate collaboration. Such approaches rely either on lock mechanisms~\citep{kelly2017collaborative} or manual conflict resolution~\citep{taentzer2010conflict}. As a consequence, they are not suitable for real-time collaboration.

\subsection{Real-time collaboration}\label{sec:background-rtcollab}

Recent studies on collaborative model-driven software engineering show a strong shift towards real-time collaboration \citep{david2021collaborative, franzago2018collaborative}.
The main challenge in such a shift is ensuring appropriate convergence of distributed replicas, while still guaranteeing timely execution \citep{sun1998achieving}.
Convergence of replicas is ensured by the consistency model a distributed setting chooses. Strict consistency is a theoretical model for guaranteeing deterministic consistency by the total order of change updates that are exchanged instantaneously. However, due to its limited usability, various relaxations have been provided, such as sequential consistency \citep{lamport1978time}, causal consistency \citep{du2014gentlerain}, and eventual consistency \citep{vogels2009eventually}. Strong eventual consistency (SEC)
\citep{preguicca2009commutative}
augments the liveness property of eventual consistency (all change updates will be observed eventually) with a safety guarantee: two nodes that have received the same set of change updates will be in the same state, regardless of the order of updates. SEC is an efficient resolution of the CAP theorem by \citet{brewer2012cap}, suggesting that distributed systems cannot provide more than two out of the three properties of strong consistency, availability, and partition tolerance. SEC removes the problem of conflict resolution on local replicas by introducing rules to ensure a unique outcome for concurrent changes, deterministically resolving any conflict. There is no need for a consensus or synchronization since any kind of change is allowed and conflicts are removed altogether. As such, this consistency model is especially appropriate for real-time collaboration.
Nevertheless, SEC may be challenging or even impossible to implement for certain data types.

\subsection{Conflict-free replicated data types}\label{sec:background-crdt}
CRDT eliminate conflicts between the distributed stakeholders' operations; thus avoiding the complexity of conflict resolution and roll-back. As a result, CRDT exhibit promising fault tolerance and reliability properties.

CRDT come in two flavors. State-based CRDT are structured in a way that they adhere to a monotonic semi-lattice. Shapiro et al.~\cite{shapiro2011conflict} show that state-based objects that satisfy this property are SEC; hence, they converge to a consistent state.
Operation-based CRDT require that concurrent operations are commutative. Independent from the order of the received change updates, the state of the local copy converges to the same state. To achieve such a behavior, the supporting communication protocol has to provide a causal ordering mechanism, such as global timestamps.
In our approach, we have opted for the operation-based CRDT scheme because of its reduced costs when exchanging model changes between replicas.

Operation-based model representation \citep{lenoir2011operation} has been proposed for reasoning about streams of model operations. The C-Praxis approach \citep{michaux2011semantically} defines six CRUD model operations and defines how these operations interact. The CRDT of \lowkey{} are geared toward the more complex semantics of graphs and, thus, they provide a superset of these operations. Additionally, \lowkey{} enables working with multigraphs, hypergraphs, and disconnected graphs, allowing for more flexibility in modeling. Traditional operation-based approaches, such as C-Praxis are built on MOF.
Consequently, they fall short of supporting arbitrary meta-levels of modeling.

\subsubsection{The Last-Writer-Wins (LWW) paradigm}

To ensure the convergence of local replicas, their differences have to be resolved in an automated fashion. Such a resolution mechanism can be implemented either in the application or at the data level \citep{meiklejohn2015lasp}. The LWW paradigm \citep{johnson1975maintenance} has been widely adopted as a data-level implementation of operation-based conflict resolution \citep{thomas1979majority,roh2011distributed}.
Conflicting operations are resolved using a global ordering operator, e.g., a timestamp.
Given two changes, it is the more recent one that will prevail \citep{shapiro2011comprehensive}.
To avoid potential data loss, each change update is stored locally and the resolution of conflicts is carried out by the local replica.

\begin{figure}
    \centering
    \includegraphics[width=\linewidth]{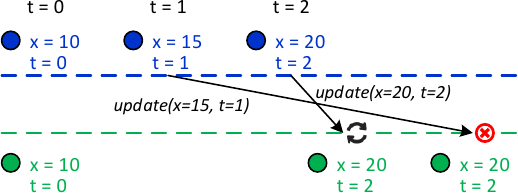}
    \caption{Total order of updates in the LWW paradigm.}
    \label{fig:lww-total-order-example}
\end{figure}

\figref{fig:lww-total-order-example} shows an example resolution scenario under LWW. User A (top blue) and User B (bottom green) initially have their local replicas in consistent states: the value of $x$ and its timestamp $t$. At $t=1$, User A executes the update $x=15$ on his local copy. A message with this updated value and the timestamp is sent to User B. However, before the message arrives, User A executes another update: $x=20$, at time $t=2$. Again, an update message is composed and sent to User B. Due to network delays, the second update arrives to User B earlier than the first. Upon receiving the update message, User B will reconcile this new value with his local replica. As it stands, User B has $x=10$ timestamped with $t=0$; and an update that says $x=20$ timestamped with $t=2$. Under the LWW paradigm, the latter value prevails, due to the more recent timestamp. Eventually, the first message arrives. User B has $x=20$ timestamped with $t=2$; and an update of $x=15$ timestamped with $t=1$. Under the LWW paradigm, the former value should prevail. Thus, the update is not performed on the local copy. Eventually, the replicas are in consistent states: $x=20$.

Despite the omitted update after receiving $x=15,t=1$, the message can still be stored in User B's local replicas. It is the responsibility of the CRDT implementation to decide whether to store outdated data or not. Some use-cases might require such behavior. However, the performance of CRDT is proportional with the data they store \citep{sun2020real}, and often requires implementing complex garbage collection mechanisms \citep{bauwens2019memory}.

The LWW paradigm satisfies the requirements for real-time collaboration defined by Sun et al.~\cite{sun1998achieving}: (i) convergence, i.e., every stakeholder's local data must exhibit the same state after updates have been applied; (ii) user intention preservation, i.e., the original user's intention must be preserved; (iii) causality preservation, i.e., updates must be ordered in the same causal way by each stakeholder; and (iv) timely execution, i.e., operations must propagate and the system must reconcile within a deadline that provides a smooth user experience.

\subsubsection{CRDT frameworks}
Yjs\footnote{\url{https://github.com/yjs/yjs}\label{fn:yjs}} is an open-source framework for peer-to-peer shared editing of structured data, such as rich-text, or XML.
Operations are stored in a linked list,
resulting in a total order of operations, thus implementing CRDT.
Teletype\footnote{\url{https://github.com/atom/teletype-crdt}\label{fn:teletype}} provides string-wise sequence CRDT for peer-to-peer collaborative editing in
the Teletype for Atom\footnote{\url{https://teletype.atom.io/}\label{fn:teletype-for-atom}}
cooperative source code development environment.
AutoCouch~\citep{grosch2020autocouch} is a JSON framework combining the benefits of the Automerge CRDT library\footnote{\url{https://github.com/automerge/automerge}} and the CouchDB\footnote{\url{http://couchdb.apache.org/}} database engine. Conflict-free JSON documents are replicated both on the server side and client side, while ensuring a responsive real-time user experience for web-based applications.

These frameworks are similar to \lowkey{} in their aim to augment engineering tools with a CRDT-based collaboration service. However, they are primarily geared toward linear data types, whereas the CRDT layer of \lowkey{} is primarily aimed at supporting a wide range of modeling scenarios.
Currently, no other modeling framework implements support for graph CRDT as \textit{first-class citizens}. Shapiro et al.~\cite{shapiro2011conflict} formalize graph CRDT but provide no implementation. SyncMeta \citep{derntl2015near} (built on top of Yjs) provides support for modeling graph-like data structures. However, graphs are emulated by linked lists and string comparisons at the CRDT level.

\subsection{Multi-level modeling}

Multi-level modeling (MLM) is a paradigm aiming to improve the flexibility of modeling architectures by introducing arbitrary meta-levels~\cite{atkinson2001essence}.
A formal framework of modeling at arbitrary meta-levels has been developed by \citet{atkinson2000meta}, introducing the core concepts of deep instantiation \citep{atkinson2002rearchitecting} and deep characterization \citep{atkinson2008reducing}. Deep instantiation extends the traditional two-level instantiation and allows classes to be instantiated transitively. This is achieved by the notion of potency that defines how many levels of instantiation the class supports. Deep characterization allows meta-types to influence the characteristics of their instances beyond those in the level immediately below. Our framework embraces these concepts to provide collaborative support in a highly generalized fashion. As shown by \citet{atkinson2002rearchitecting}, deep metamodeling can naturally accommodate traditional modeling frameworks built on shallow instantiation, such as UML and EMF. Consequently, our framework is a good fit with modeling tools supporting UML and EMF modeling, but lacking collaborative features.

The feasibility of multi-level modeling has been demonstrated in tools such as MetaDepth \citep{delara2010deep} and Melanee \citep{atkinson2016flexible}. However, as pointed out recently by \citet{kuhne2022multi}, the divergence of MLM techniques has led to numerous different interpretations of the nature of levels, their purpose, the appropriateness of implementation techniques, etc. Often, the dual ontological and linguistic instantiation of the Orthogonal Classification Architecture (OCA)~\cite{atkinson2009flexible} allows elements to be typed by linguistic concepts and domain concepts at the same time~\cite{alvarez2001mapping,atkinson2002rearchitecting}. In this paper, we focus on a third orthogonal dimension, the \textit{physical dimension} that deals with the physical representation of concepts. The physical–logical distinction was first introduced by \citet{atkinson2002rearchitecting}. Specifically, we build on the adapted version of the physical metamodel defined by \citet{van2014multi}. The OCA usually omits the physical dimension, lumping it into the linguistic dimension and relying on its physical layer for representation and serialization (e.g., XMI in MOF).
However, in collaboration techniques operating at the data level---such as CRDT---the management of consistency is achieved at the physical level of models. This necessitates the separation of the physical and linguistic dimensions. The explicit physical dimension enables ensuring physical consistency while allowing linguistic inconsistencies. Temporary linguistic inconsistencies are often desirable, especially when complex modeling operations are used, or when atomic modeling steps do not result in a linguistically valid model (see Scenario 2 in \secref{sec:elaboration-collaboration}).
For the sake of conciseness, we choose to treat the physical dimension of \citet{atkinson2002rearchitecting} and the linguistic dimension of the OCA explicitly, and treat the ontological dimension implicitly. However, because CRDT operates at the data level, our framework is able to treat both linguistic and ontological conformance.

%% file: sections/framework.tex
\section{A framework for real-time collaborative metamodeling}\label{sec:framework}

We present the different components of \lowkey{}, our real-time collaborative modeling framework.

\subsection{Architecture}\label{sec:framework-architecture}

\begin{figure*}[!ht]
    \centering
    \includegraphics[width=\linewidth]{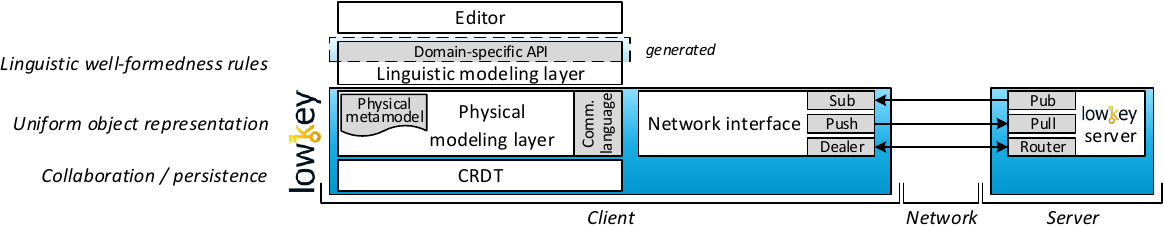}
    \caption{Overview of the architecture and the roles of each layer}
    \label{fig:architecture-highlevel}
\end{figure*}

\begin{figure*}[htb]
    \centering
    \includegraphics[width=0.8\linewidth]{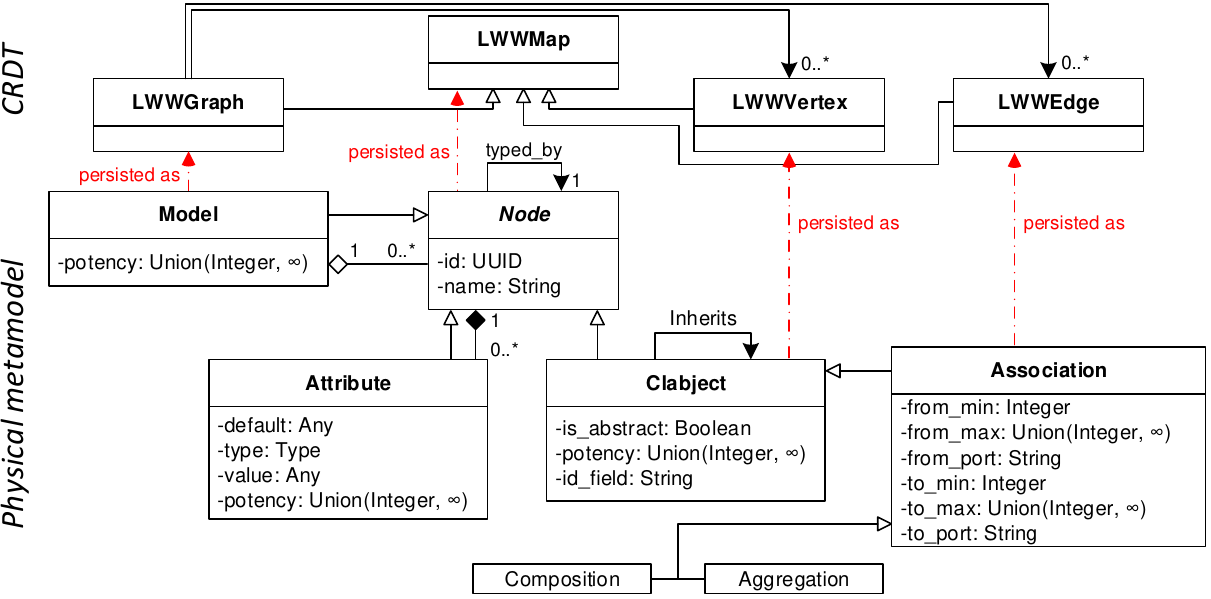}
    \caption{The Physical metamodel and CRDT layers}
    \label{fig:physical-metamodel-crdt}
\end{figure*}

\figref{fig:architecture-highlevel} outlines the architecture of the framework.
In a typical modeling setting, users are provided with (domain-specific) \textit{Editors} that enable interacting with the models of the \textit{linguistic modeling layer}, through a generated \textit{domain-specific API}. The \textit{linguistic modeling layer} enforces well-formedness rules defined by the static semantics of the language. Each element at the \textit{linguistic modeling layer} conforms to a \textit{physical metamodel}, as described by Van Mierlo et al.~\cite{van2014multi}. The \textit{physical modeling layer} is responsible for the uniform representation of objects of the linguistic models and metamodels. Instances of the physical metamodel are serialized and propagated to the collaborating stakeholders through their \textit{network interface} which communicates with the server. Other clients receive these instances from the server and merge them into the local working data represented in CRDT. The \textit{command language} of the \textit{physical modeling layer} enables a uniform treatment of local updates (from the \textit{linguistic modeling layer}), and remote updates (from the \textit{network interface}).

\subsection{Linguistic modeling layer}\label{sec:framework-linguistic}
The linguistic modeling layer provides mechanisms for modeling at arbitrary levels of abstraction. As a consequence, this layer contains every domain model and its metamodels the Editor manipulates. To this end, a domain-specific API is generated for the metamodels. In the illustrative case, the \textit{MindMap} class and its instance(s) are situated at this modeling layer; and methods for CRUD operations are generated.

The main responsibility of this layer is to enforce the language-specific well-formedness rules defined by the meta-model and static semantics. Typical examples of language-specific well-formedness rules include multiplicities, transitive containment by compositions, and uniqueness of marker names.

\subsection{Physical modeling layer}\label{sec:framework-physical}

The main responsibility of the physical modeling layer is to provide a metamodel that every linguistic concept can conform to, regardless of which linguistic meta-level they are situated at. For example, the physical metamodel has to accommodate both the \textit{MindMap} class; its instance \textit{mindmap\_0}; and the \textit{Class} class the \textit{MindMap} class corresponds to. In addition, the physical modeling layer may impose language-independent well-formedness rules, such as each model having to exhibit a graph structure.

\subsubsection{Physical metamodel}

We adopt the physical metamodel from previous work \citep{van2014multi}, as shown at the bottom of \figref{fig:physical-metamodel-crdt}.

\textit{Node} is the foundational concept of the metamodel, which can be organized into \textit{Model}s. Models, in turn, are \textit{Node}s themselves, allowing for the hierarchical composition of \textit{Model}s.

A consequence of metamodeling---and multi-level\\ modeling in particular---is that instantiable model elements can play the role of both instances and types~\citep{atkinson2000meta}. To accommodate this dichotomy, \textit{Clabject}s serve as the physical metatype for every linguistic class and instance. For example, in the mindmap metamodel, both the \textit{CentralTopic} class and its instance(s) are mapped onto the \textit{Clabject} physical type.

\textit{Association}s link \textit{Clabject}s to each other. The \textit{Association} inherits from the \textit{Clabject}, and transitively from the \textit{Node}. Due to the latter, \textit{Association}s can be typed by other \textit{Nodes}. Due to the former, \textit{Association}s can be abstract and link other \textit{Association}s. We allow this flexibility to accommodate the models of various modeling frameworks and formalisms that typically implement a subset of these modeling options. UML, for example, restricts \textit{Association}s from being abstract, and only allows linking \textit{Clabject}s.
\textit{Composition} and \textit{Aggregation} are specialized types of \textit{Association} with conventional semantics.

\textit{Attribute}s store information of specific \textit{Node}s. Specifically, in \textit{Attribute}s, the \textit{type} property maintains information about the linguistic type the \textit{Attribute} corresponds to. For example, storing textual information (such as the \textit{name} of a \textit{Topic} in the illustrative case) in the Eclipse Modeling Framework (EMF) would mean the type of the \textit{Attribute} is \textit{EString}. \textit{Attribute}s are specializations of the \textit{Node} type. The inherited \textit{typed\_by} relationship provides a mechanism for typing an \textit{Attribute} at the physical level.

\subsubsection{Need for an explicit physical metamodel}
Instantiation mechanisms of current modeling frameworks typically consider two levels: classes and their instances (objects). Although this covers the majority of practical use cases, some scenarios might require multiple levels of type-instance relationships \citep{delara2014when}. Such a type-instance hierarchy enables various beneficial mechanisms, such as deep instantiation and deep characterization \citep{atkinson2002rearchitecting}. The role of a physical metamodel is to represent the objects of such a type-instance hierarchy uniformly, irrespective of what metalevel a specific object is situated at; and to provide services such as the serialization of these objects. By this separation of concerns, objects of the type-instance hierarchy become independent of their physical representation, and their roles are determined by purely linguistic concepts, such as being metatypes or instances of each other. Therefore, we refer to this level of objects as the linguistic metamodel.

The physical metamodel of traditional modeling frameworks is typically coupled with the topmost meta-level of their linguistic metamodels. For example, Ecore serves as the core metamodel of EMF and also defines the rules governing the serialization of EMF models into XMI files.

By providing mechanisms for multi-level metamodeling, and clearly separating the physical metamodel from the linguistic metamodel, our framework is able to (i) support advanced metamodeling scenarios; and (ii) accommodate traditional modeling frameworks, for example, by restricting the flexibility of the physical metamodel. Similar avenues have been explored by multiple modern modeling frameworks, such as Melanee \citep{atkinson2016flexible}, metaDepth \citep{delara2010deep}, and the Modelverse \citep{van2017modelverse}.

\begin{lstlisting}[float=*, language={Command}, caption = {Command language for interacting with the physical metamodel}, label = {lst:commands}]
CREATE -name {name} -typedBy {type} [-attributeName {value}]*
LINK -from {fromClabject}.{associationName} -to {toClabject} [-attributeName {value}]*
UPDATE (-name {name} | -id {id}) [-attributeName {newValue}]*
DELETE (-name {name} | -id {id})
\end{lstlisting}

\subsubsection{Command language} The physical metamodel can be accessed and interacted with through a command language. The brief definition of the language is shown in \lstref{lst:commands}. The command language defines four CRUD operations. Therefore, integrating an editor with the physical metamodel requires the appropriate facility that translates domain-specific modeling operations to the operations of the command language. 
We have opted for providing an external textual DSL because such languages are more trivially serialized and propagated through standard network protocols than binary data.

\subsection{CRDT layer}\label{sec:framework-crdt}

The CRDT layer is responsible for persisting instances of the physical metamodel. As shown in~\figref{fig:physical-metamodel-crdt}, each element of the physical metamodel is associated with exactly one CRDT. At run-time, as the classes of the physical metamodel get instantiated, a corresponding CRDT is instantiated as well. The model element maintains a reference to its persisting CRDT instance during execution.
Our CRDT implementations follow the LWW paradigm and have been mainly implemented by following the specifications outlined by Shapiro et al.~\cite{shapiro2011comprehensive}. In the following, we briefly elaborate on these types.

\subsubsection{Timestamps for total order}\label{sec:timestamps}
To ensure the commutativity of operation-based CRDT, an operator for total order is required. The most natural choice in a distributed setting is a global timestamping mechanism, as it makes the fewest assumptions about the system.
\lowkey{} uses the Unix epoch time in nanoseconds to timestamp model updates with. As an alternative, Lamport clocks~\cite{lamport1978time} or vector clocks~\cite{singhal1992efficient} can be used as suggested by Shapiro et al.~\cite{shapiro2011comprehensive}. Timestamped model updates are subsequently forwarded to the collaborating stakeholders. The global nature of the timestamp ensures that each local replica sorts the updates in the same order, ensuring the safety property of SEC (c.f. \secref{sec:background-crdt}).
The appropriate granularity of
timestamps is paramount in supporting complex CRDT. Consider the underlying data structure in the illustrative example in \figref{fig:lww-total-order-example} being a graph, in which vertices represent entities with associated attributes. Assigning a timestamp to the whole graph would not allow independent changes at finer-grained levels, such as vertices. Thus, we assign a timestamp to each instance of the elements of the physical metamodel,
and update it on each CRUD operation.

Another useful operator could be the priority of messages, e.g., to define hierarchical stakeholder roles in which higher ranked roles can overwrite the changes of lower-ranked roles. An important difference between timestamps and priorities is the level of granularity. Timestamps (e.g., at the level of nanoseconds) provide better chances to unambiguously order two operations, as compared to priorities (e.g., assigned from a range between 1--5). Therefore, the ordering operator should be carefully designed to ensure the total order. Furthermore, compound ordering operators can be used as well. For example, the shortcomings of the priority operator can be circumvented by combining it with timestamps. Such compound operators are best defined by linking them through the absorption identity over the lattice of change operations~\citep{davey2002introduction} to ensure a valid and sound composition.

\subsubsection{LWWRegister}
The \lwwregister{} is the simplest CRDT in \lowkey{}, containing a single value. To implement LWW semantics, its value \textit{v} is equipped with a timestamp \textit{t}. Thus, an \lwwregister{} $r$ is defined as $r = (v, t)$.

\operation{update}
The \textit{update} operation permits to modify the value of the register. Given a new timestamped value $(v', t')$, the \textit{update} operation on register $r = (v, t)$ is defined as $r := (v', t')$ iff $t' > t$, and \texttt{NOP} otherwise.

\subsubsection{LWWSet}
The \lwwset{} $S$ contains an arbitrary number of timestamped values $(v, t)$ under set semantics. That is, for $S = \langle (v,t) \rangle$, it holds that $\forall s_i=(v_i, t_i),\, s_j=(v_j, t_j) \in S: v_i \neq v_j$. We have implemented the \lwwset{} as an LWW-element-Set \citep{shapiro2011comprehensive}. LWW-element-Sets are composed of the \textit{add-set} $A$ and the \textit{remove-set} ($R$). The two sets are disjoint with the \textit{add-set} containing values added to $S$, and the \textit{remove-set} containing values removed from $S$. This structure ensures the commutativity of \textit{add} and \textit{remove} operations: \textit{add(v, t)}~$\circ$~\textit{remove(v, t')}~$\equiv$~\textit{remove(v, t')}~$\circ$~\textit{add(v, t)}.

\operation{lookup}
The \textit{lookup} operation indicates if an element is in the set.
A value is considered to be in the \lwwset{} iff it can be found in the \textit{add-set}, and it cannot be found in the \textit{remove-set} with a higher timestamp. It is defined as $\textit{lookup}: v \to \mathbb{B}$, where $\textit{lookup}(v)$ evaluates to \textit{true} if $\exists v,t\,\nexists t': (v,t) \in A \wedge (v,t') \in R, t'>t$.

\operation{add}
The \textit{add} operation inserts new values in the \textit{add-set}. It is defined as $\textit{add}(v,t): S.A \to S.A \cup \{(v, t)\}$.

\operation{remove}
The \textit{remove} operation deletes values from the set by adding them to the \textit{remove-set}. It is defined as
\textit{remove(v, t):} $S.R \to S.R \cup \{(v,t)\}$.

\subsubsection{LWWMap}\label{sec:framework-crdt-map}
The \lwwmap{} is defined as an extension of the \lwwset{}. The data is stored as a key/value pair $((k, v), t)$.

\operation{lookup, query} The \textit{lookup} operation is modified so that it looks up the key instead of the value. That is,
$\textit{lookup}: k \to \mathbb{B}$, where $\textit{lookup}(k)$ evaluates to \textit{true} iff $\exists k,t\,\nexists t': ((k,v),t) \in A \wedge ((k,v),t') \in R, t'>t$. The $\textit{query}(k)$ method returns the value for a key $k$.

\operation{add, remove} Analogously to the \textit{add} and \textit{remove} of the \lwwset{}: \textit{add((k, v), t):} $S.A \to S.A \cup \{((k, v), t)\}$; and \textit{remove((k, v), t):} $S.R \to S.R \cup \{((k,v),t)\}$.

\operation{update} This operation is the only substantial difference the \lwwmap{} introduces to the \lwwset{}. Updating a key $((k, v), t)$ entry with a value $v'$ timestamped with $t'$ is defined as $\textit{update}(k,v,v',t,t') = add((k,v'),t')~\circ$ $\textit{remove}((k,v),t'-\epsilon)$.
First, the entry with key $k$ is removed, and the time of removal is timestamped with a value that is older than the new timestamp $t'$ by the minimal time interval $\epsilon$ the system is able to detect. In \lowkey{} $\epsilon$  = 1 ns. Subsequently, the entry with key $k$ and the new value $v'$ is added with the timestamp $t'$.

\subsubsection{LWWGraph}
The \lwwgraph{} \textit{G} is the extension of the \lwwmap{} that enables persisting attributes of the graph. This mechanism is used for storing the vertices and edges of the graph. Vertices are stored in an \lwwset{}, denoted by $V$. Each element of $V$ is an \lwwvertex{}. To enable storing attributes and metadata of vertices, each \lwwvertex{} is an extension of the \lwwmap{}. Analogously, the edge set $E$ is an \lwwset, containing \lwwedge{} instances. Formally: $G=(V, E)$, where
\begin{itemize}
    \item $V, E \vdash$ \lwwset{};
    \item $V = \langle(v,t)\rangle$, where $v \vdash$ \lwwvertex{} and $t$ is a timestamp;
    \item $E = \langle(e,t)\rangle$, where $e \vdash$ \lwwedge{} and $t$ is a timestamp;
    \item $\forall e \in E: e.\textit{query}(\textit{``source''}), e.\textit{query}(\textit{``target''}) \in V$ (denoted \textit{e.source} and \textit{e.target}).
    \item \lwwgraph{}, \lwwvertex{}, \lwwedge{} $\vdash$ \lwwmap{}.\\
\end{itemize}

Due to the invariant property of $E \subseteq V \times V$, operations on $V$ and $E$ are not independent. Shapiro et al.~\cite{shapiro2011comprehensive} suggest multiple ways to manage this issue. In \lowkey{}, we chose prioritizing the \textit{removeVertex} operation to ensure CRDT behavior. That is, the operation is only allowed to be executed if it does not leave a dangling edge behind.
\paragraph{lookup and query} Since both the \lwwvertex{} and the \lwwedge{} extend the \lwwmap{}, their lookup and query methods are identical to the ones discussed in \secref{sec:framework-crdt-map}.
\paragraph{addEdge, addVertex} These operations reuse the API of the \lwwset{} directly.
Adding an edge is achieved by adding the edge to the add-set $A$ of the edge set $E$ of graph $G$.
Thus, $\textit{addEdge}(e,t): G.E.A \to G.E.A \cup \{(e, t)\}$; and $\textit{addVertex}(v,t): G.V.A \to G.V.A \cup \{(v, t)\}$.
\paragraph{removeEdge, removeVertex} These operations reuse the API of the \lwwset{} directly.
Removing an edge is achieved by adding the edge to the remove-set $R$ of the edge set $E$ of graph $G$.
Thus, $\textit{removeEdge}(e,t): G.E.R \to G.E.R \cup \{(e, t)\}$; and $\textit{removeVertex}(v,t): G.V.R \to G.V.R \cup \{(v, t)\}$.
Furthermore, to ensure the CRDT behavior: $\textit{removeVertex}(v,t) \Rightarrow {\not\exists}$~$e \in E:$ \textit{e.source=v} $\vee$ \textit{e.target=v}.

Additional methods of the \lwwgraph{} are defined for querying various properties of the graph; and for adding and removing vertices and edges by name and identifier. Additional methods of the \lwwedge{} and \lwwvertex{} types are defined, \eg for querying incoming and outgoing edges of a vertex, cascading the removal of dangling edges upon a vertex removal, and assigning direction to edges.

\subsection{Network architecture}\label{sec:framework-network}

\lowkey{} follows a client-server network architecture with multiple clients connecting to the same server, as outlined already in \secref{sec:framework-architecture}.
The network architecture is built on top of the ZeroMQ\footnote{\url{https://zeromq.org/}} asynchronous messaging library. It offers efficient scalability and latency properties; thus it aligns well with the requirements of real-time collaboration.

\subsubsection{Server and Client components}
As shown in \figref{fig:architecture-highlevel}, our framework provides two network components for connecting remote collaborating stakeholders: the \textit{server} and the \textit{client}.

The server component is responsible for two tasks: (i) collecting from, and distributing updates among clients; and (ii) providing newly joined clients with the snapshot of the system so they have an initial local replica.

The client component is responsible for providing networking capabilities to modeling tools and editors. The \textit{Client interface} is accessible from the API of the framework and it is the tool builder's responsibility to properly implement its required functionality. Specifically, the \textit{Client interface} is properly implemented by defining the action the \textit{Sub} socket executes periodically during its polling loop.

\subsubsection{Communication patterns}
Different responsibilities of the server are implemented with different communication patterns, as summarized in \figref{fig:architecture-highlevel}.

\paragraph{Exchanging updates.}
To receive updates, a client must first \textit{subscribe} to the updates by connecting to the \textit{Pub} (publisher) socket of the server with its own \textit{Sub} (subscriber) socket. The Pub-Sub pattern establishes a one-way asynchronous communication channel with the client receiving and processing messages in a polling loop. The Pub socket broadcasts messages to every client connected to the server.
To send updates, a client must first connect to the \textit{Pull} socket of the server with its \textit{Push} socket. Similar to the Pub-Sub pattern, the Pull-Push pattern establishes a one-way connection. Note that the \textit{Push} socket is geared towards supporting pipelining mechanisms, hence it does not broadcast messages.

\paragraph{Distributing system snapshots.}
Upon connecting to the server, the client needs to obtain a snapshot of the system. This is achieved by connecting to the \textit{Router} socket of the server via the \textit{Dealer} socket of the client. The Router-Dealer pattern implements a request-reply mechanism with both ends acting asynchronously.
As opposed to the patterns used for exchanging updates, this pattern does not maintain a connection after the request-reply pair of messages has been exchanged.

To provide a snapshot to its clients, the server is equipped with a memory that stores the history of previously exchanged messages (model updates). Once the newly connected client requests the snapshot, the history of messages is replayed to it. The state of the system is then built up locally by the client. We opted for this mechanism to keep it aligned with our choice of operation-based CRDT semantics (\secref{sec:background-crdt}), and to shift the workload to the client instead of the server. Since operation-based and state-based CRDT are able to emulate each other, it would be possible to communicate the whole state at once, but this would come at the price of increased network traffic.

\subsubsection{Update messages} Every update message has the following signature:\\
$\langle \textit{clientId, command, timestamp} \rangle$, where the \textit{clientId} is a UUID assigned to the client upon its creation, used for avoiding double delivery problems; \textit{command} corresponds to the command language of the physical metamodel shown in \lstref{lst:commands}; and \textit{timestamp} is the timestamp of creation, as discussed in \secref{sec:timestamps}. Messages are serialized as text and sent through the TCP/IP stack.

Upon reception, the command encapsulated in the update message is executed on the local replica. In the current implementation of \lowkey{}, the scope of the change is the root model in the modeling session. Advanced modeling scenarios might necessitate more complex scope management. For example, multi-view modeling requires support for namespaces in order to allow independent views to define syntactically similar or identical concepts~\cite{corley2016cloud}. Such features are left for future work.

%% file: sections/elaboration.tex
\section{Feasibility evaluation}\label{sec:elaboration}

Based on our illustrative case, we evaluate the feasibility of \lowkey{} from two points of view. In both cases, our objective is to assess whether \lowkey{} can accommodate the challenges outlined in the case study. First, in \secref{sec:elaboration-metamodel}, we provide a language engineer's view by presenting the process of metamodeling in \lowkey{}. Then, in~\secref{sec:elaboration-collaboration}, we provide a domain expert's view by demonstrating four collaborative modeling scenarios.

\paragraph{Study setup.} We conduct the evaluation on three desktop machines with Python 3.7 and the required libraries installed on them. We deploy the \lowkey{} framework as an editable local package. First, in \secref{sec:elaboration-metamodel}, we model the domain concepts of the Mind map editor on one machine and generate a domain-specific API the other two machines will use during collaboration. Second, in \secref{sec:elaboration-collaboration}, we deploy this domain-specific API to two machines along with the Mind map editor and conduct our experiments guided by the requirements formulated in \secref{sec:case}.

\subsection{Language engineer's view: metamodeling in \lowkey{}}\label{sec:elaboration-metamodel}

\begin{figure*}[t]
    \centering
    \includegraphics[width=\textwidth]{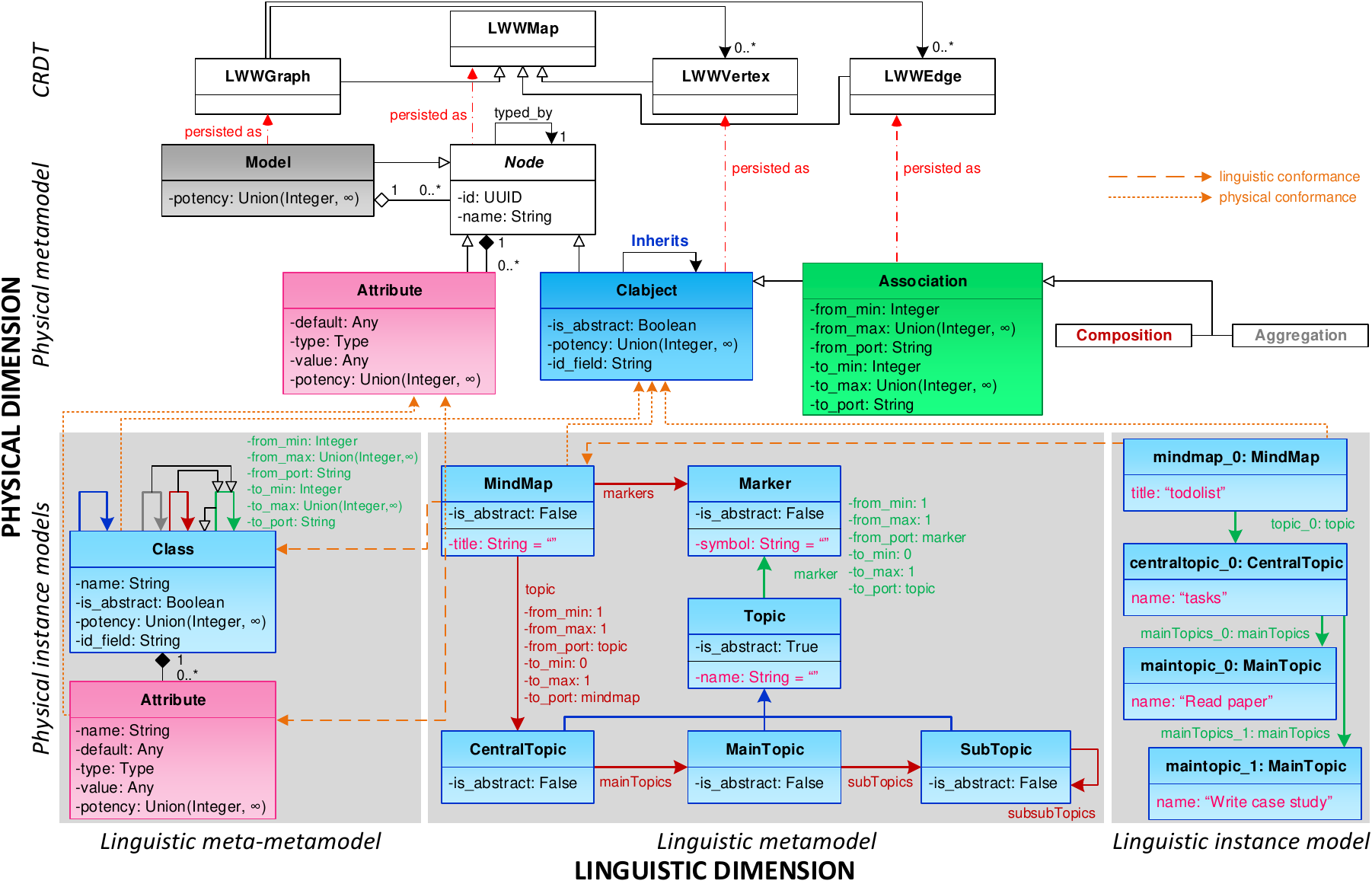}
    \caption{The three-level hierarchy of linguistic models corresponding to the physical metamodel in the mindmap case}
    \label{fig:elaboration-metamodel-case}
\end{figure*}

\figref{fig:elaboration-metamodel-case} shows how the metamodel of the illustrative case (\figref{fig:mindmapmm}) is defined within the \lowkey{} framework in relation to the physical model (\figref{fig:physical-metamodel-crdt}).
The \textit{linguistic metamodel} represents the same domain as the metamodel in \figref{fig:mindmapmm}, but instead of using UML to express it, here, we use the \textit{physical metamodel} of the \lowkey{} framework.\footnote{We remark, that multi-level modeling approaches traditionally rely on the orthogonal linguistic and \textit{ontological} dimensions, and do not consider the physical dimension. In this paper, we only consider the linguistic and physical dimensions to allow an easier discussion. Our approach can be safely extended with ontological aspects, similarly to the work of \citep{van2014multi}.}
The color coding shows how every concept \textit{physically conforms} to the \textit{Clabject}. This is due to the \textit{linguistic metamodel} being an \textit{instance} of the \textit{physical metamodel}. In addition, the \textit{linguistic metamodel} \textit{linguistically conforms} to the \textit{linguistic meta-metamodel}. The latter is another instance of the \textit{physical metamodel}; thus, its elements \textit{physically conform} to the specific elements of the \textit{physical metamodel} (shown by color-coding). Specific Mind map instances are created in the \textit{linguistic instance model} that \textit{linguistically conforms} to the \textit{linguistic metamodel}, and \textit{physically conforms} to the \textit{physical metamodel}.
As emphasized by the arrow notation, the \textit{mindmap_0} object is a linguistic instance of the \textit{MindMap} class; which is, in turn, a linguistic instance of the \textit{Class} class. Each of these are physical instances of the \textit{Clabject} class. Similarly, the title of the mindmap \textit{"todolist"} is a linguistic instance of \textit{title: String = ""}; which is, in turn, a linguistic instance of the \textit{Attribute} class.

Thanks to the clear physical conformance relationships, every element at the \textit{physical instance} level will be persisted as a CRDT, irrespectively of their linguistic meta-level. Therefore, stakeholders can safely collaborate by editing any of the three \textit{linguistic metamodels}.

\paragraph*{Generating a domain-specific API.} To enable interacting with the \textit{linguistic metamodel}, an API has to be produced.
This process is fully automated with a template-based code generator that produces a Python class for every Clabject, and generates the appropriate accessors (\eg get, set, add, remove) for attributes and references.
\lstref{lst:api} shows the method signatures of the Python code generated for the \textit{MindMap} class.\footnote{The full example is available at \url{https://github.com/geodes-sms/lowkey}.\label{fn:example-gh}}

\input{src/MindMap.py}
\vspace{-10px}

\subsection{Domain expert's view: collaboration in \lowkey{}}\label{sec:elaboration-collaboration}

\begin{lstlisting}[float=*, language={Command}, caption = {MindMap DSL of the Editor}, label = {lst:commands-editor}]
READ -- Returns the mindmap model in a readable form
OBJECTS -- Lists every object in the local session
CREATE {type} {name} -- Creates an instance with name of the domain-specific type
LINK {source}.{port} TO {target} -- Links object target to object source via port
UPDATE {name} {attribute} {newValue} -- Updates attribute with name to newValue
DELETE {name} -- Deletes object name
\end{lstlisting}

\begin{figure*}[htb!]
    \begin{subfigure}{\textwidth}
        \centering
        \includegraphics[width=\textwidth]{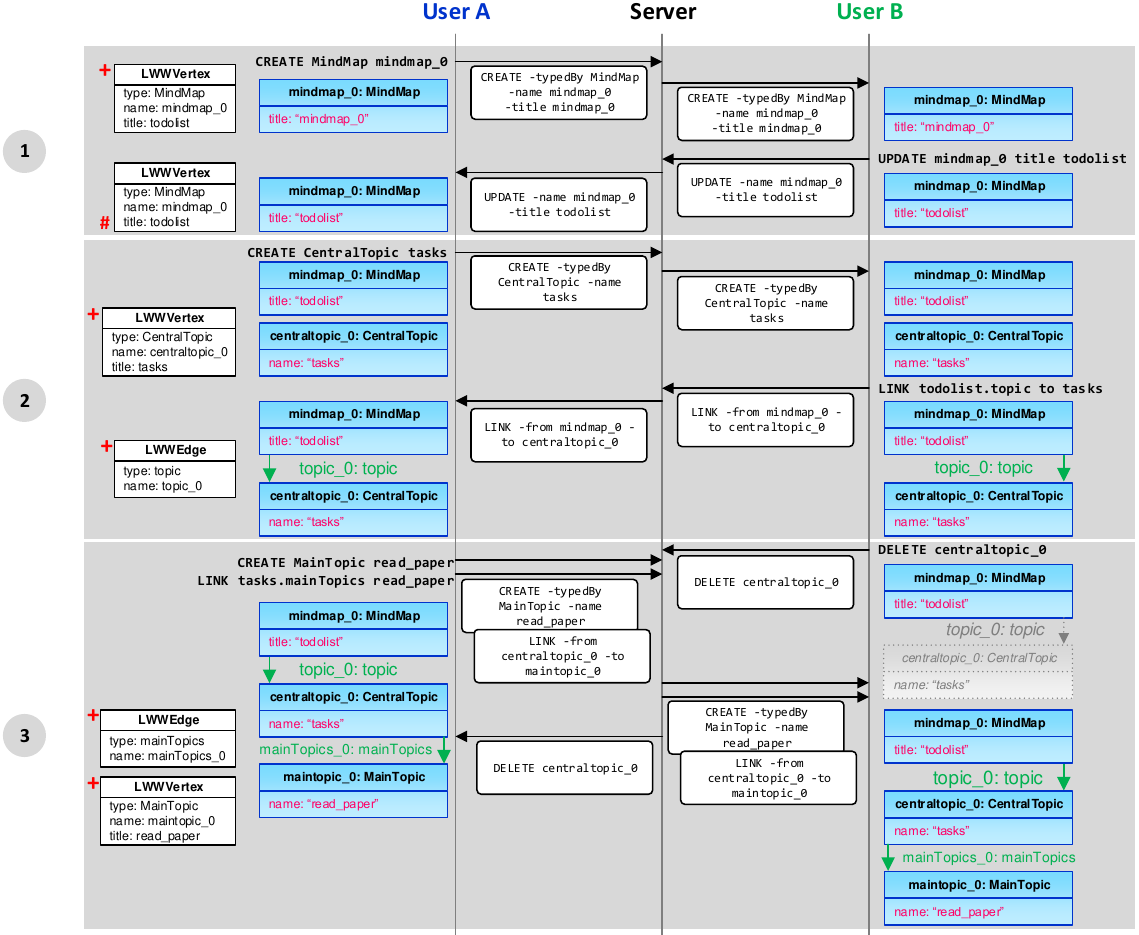}
        \caption{Changes in the CRDT of Client A in Scenarios 1--3}
        \vspace{10px}
        \label{fig:elaboration-process-a}
    \end{subfigure}
    \begin{subfigure}{\textwidth}
        \centering
        \includegraphics[width=\textwidth]{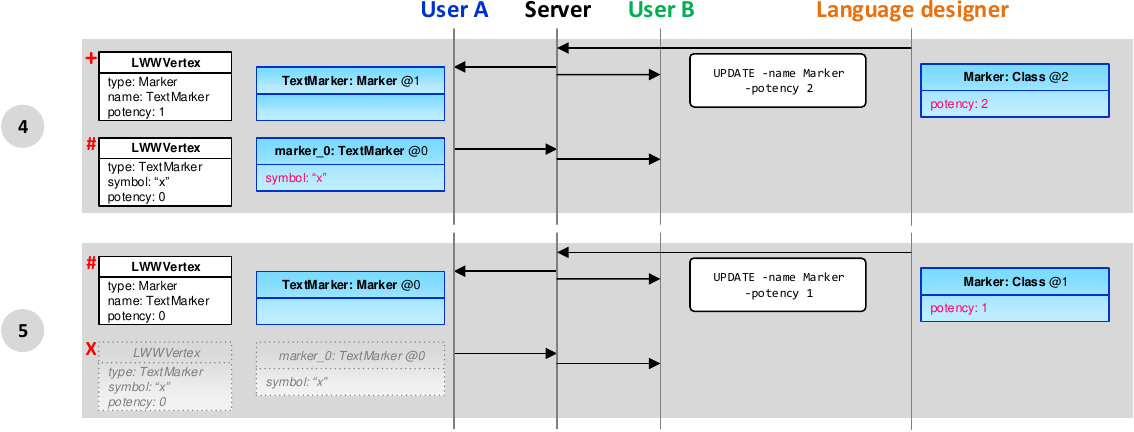}
        \caption{Changes in the CRDT of Client A in Scenarios 4--5}
        \label{fig:elaboration-process-b}
    \end{subfigure}
    \caption{Collaboration in \lowkey{} outlining changes in the CRDT shown for Client A}
    \label{fig:elaboration-process}
\end{figure*}

After the \textit{linguistic metamodel} has been defined, the domain experts can build the \textit{linguistic instance model} in a collaborative fashion. \figref{fig:elaboration-process} illustrates this collaboration by showing the objects of the \textit{linguistic metamodel} and the CRDT at the local replicas. For space considerations, we omit the technical details, such as clients connecting to the server; and only show the CRDT for \textit{Client A}.
For testing, evaluation, and demonstration purposes, we implemented a simple Mind map editor via the command line.
The commands of the editor are shown in \lstref{lst:commands-editor}. In the following, we outline typical collaboration scenarios based on our experiments with the editor.

\subsubsection*{1. Cooperation}
\begin{itemize}[leftmargin=*]
    \item Client A creates a MindMap instance by issuing the \texttt{CREATE MindMap mindmap\_0} command. As shown in \figref{fig:architecture-highlevel}, the command is parsed into Python source code that uses the Domain-specific API (\lstref{lst:api}) of the linguistic level. The role of the linguistic level is to enforce linguistic well-formedness rules. Since the model with the single \texttt{mindmap\_0} object is well-formed, the command is further translated to the command language of the \textit{physical modeling layer} (\lstref{lst:commands}).
    
    \item The client sends this command to the Server, using their respective \textit{Push-Pull} sockets. The message is stored in the list of updates at the Server.
    
    \item At the CRDT level, an \lwwvertex{} is instantiated to store the local data at the client side. The key-value pairs of the \lwwvertex{} store every information required to reconstruct the linguistic object, i.e., (\textit{type}, \textit{MindMap}); (\textit{name}, \textit{mindmap\_0}); and (\textit{title}, \textit{mindmap\_0}). In the current implementation of \lowkey{}, an \lwwgraph{} is automatically instantiated to accommodate the \lwwvertex{} and \lwwedge{} instances created throughout the collaboration.
    
    \item The Server broadcasts the change to the collaborators, i.e., Client B. Subsequently, an object from the MindMap class is instantiated with the same ID as the one at User A's side.
    
    \item Client B changes the \textit{title} attribute of the \texttt{mindmap\_0} object by issuing the \texttt{UPDATE} command. The change is applied locally and forwarded to the Server after the required conversion to the command language. 
    Client A receives the change from the Server and updates its local replica.
    
    \item At the CRDT level, the \lwwvertex{} is updated by changing the \textit{title} from \textit{mindmap\_0} to \textit{todolist}, reflecting the changes by \textit{Client B}. As explained in \secref{sec:framework-crdt}, the previous and current values are stored in the \textit{Add-set} of the \lwwvertex{} (inherited from the \lwwmap{}). Querying the CRDT instance yields \textit{todolist} as the value with the latest timestamp.
\end{itemize}

\subsubsection*{2. Cooperation with linguistic inconsistencies}

\begin{itemize}[leftmargin=*]
    \item Client A creates a CentralTopic instance by the name \textit{tasks}. The \textit{centraltopic\_0} object is not linked to the \textit{mindmap\_0} object, violating the \texttt{1-1} relationship between the \textit{MindMap} and \textit{CentralTopic} classes, prescribed by the static semantics of the metamodel (\figref{fig:mindmapmm}). In this example, we assume a less restrictive linguistic modeling layer, in which static semantics are not enforced. The \textit{physical modeling layer}, however, enforces physical well-formedness rules, e.g., the model has to correspond to a graph, which is allowed to be disconnected. (This rule is defined by the aggregation between the \textit{Model} and \textit{Node} types in \figref{fig:physical-metamodel-crdt}.)
    The editor-level user operation, thus, is propagated to the \textit{physical modeling layer}, translated into the command language: \texttt{CREATE -typedBy CentralTopic -name tasks}. Because of the valid physical model, the message is forwareded to the server and distributed to Client B. 

    \item At the CRDT level, a new \lwwvertex{} is created to store the \textit{CentralTopic} instance.
    
    \item Client B links the CentralTopic instance to the MindMap instance. The distribution of the tasks of object creation (Step 5) and linking (Step 6) creates a truly collaborative setting, demonstrating cases in which stakeholders of specialized expertise take care of partial tasks, not necessarily resulting in a valid linguistic model on every atomic operation. Client B
    creates an association of type \textit{topic} between the two objects, named \textit{topic\_0}. The operation results in a valid linguistic model, the changes are propagated in the usual way, and eventually, the change is reflected in the replica of Client A.

    \item At the CRDT level, a new \lwwedge{} is created to store the association. The \lwwedge{} is added to the \lwwgraph{} with the previous two \lwwvertex{} as its source and destination.
\end{itemize}

\subsubsection*{3. Conflict management}
\begin{itemize}[leftmargin=*]
    \item Client A works on objects that Client B has deleted. First, Client B deletes the \textit{centraltopic\_0} object, and as a consequence, the object is marked deleted in the local replica. As explained in \secref{sec:framework-crdt}, CRDT use soft delete to be able to restore data if needed. The update message is sent to the server.
    In the meantime, Client A creates an instance of the \textit{MainTopic} class and links it to the \textit{centraltopic\_0} object via the \textit{mainTopics\_0} reference. The two update messages are propagated to the server. Client A started its changes before the updates about Client B's changes have reached him. Subsequently, both clients' updates are sent to the other client, causing inconsistency in the local replicas. However, both inconsistencies are resolved immediately by the LWW semantics of CRDT. Since Client A's updates have been created later, the timestamping mechanism will resolve the inconsistencies by (i) leaving Client A's replica intact and (ii) re-adding the \textit{centraltopic\_0} object to Client B's replica.

    \item At the CRDT level, the local replica of \textit{Client B} removes the \lwwvertex{} storing the \textit{CentralTopic} instance: the vertex is removed from the vertex set of the \lwwgraph{}. This is achieved by adding the vertex to the \textit{Remove-set} of the graph, as explained in \secref{sec:framework-crdt-map}. Before this information is propagated, \textit{Client A} performs editing operations resulting in a new \lwwvertex{} and a new \lwwedge{}. The source of the newly created \lwwedge{} is the \lwwvertex{} removed by \textit{Client B}, hence the conflict. However, due to the more recently created (and timestamped) \lwwedge{}, the \lwwvertex{} is not removed in the local replica of \textit{Client A}. After \textit{Client B} is notified about the operations of \textit{Client A}, the \lwwvertex{} storing the \textit{CentralTopic} instance is re-added to its local replica, restoring the consistency between the clients.

\end{itemize}

\subsubsection*{4. Multi-level cooperation}
\begin{itemize}[leftmargin=*]
    \item Users A and B have created multiple instances of the \textit{Marker} class and would like to categorize them into textual and graphical markers. One way to achieve this is by allowing a templating mechanism to these users so that they can create specific sub-classes of the \textit{Marker} class. These classes can be then instantiated with a specific \textit{symbol}.
    \item The Language designer increases the \textit{potency} of the \textit{Marker} class from 1 to 2. The change is distributed to the users.
    \item User A creates a new instance of the \textit{Marker} class by the name \textit{TextMarker}. The \textit{potency} of the newly created clabject is set to 1, allowing one more instantiation. Subsequently, he modifies the previous \textit{marker\_0} in a way that it is typed by the newly created \textit{TextMarker} class. The potency of \textit{marker\_0} is set to 0, preventing any further instantiation.
\end{itemize}

\subsubsection*{5. Multi-level integrity}
\begin{itemize}[leftmargin=*]
    \item After defining the \textit{TextMarker} type and creating the \textit{marker\_0} instance of it, the language designer decides to revoke this level of flexibility from the users and sets the \textit{potency} of the \textit{Marker} class to 1. The change is distributed to the users.
    \item The potency of the \textit{TextMarker} instance is now automatically set to 0 by \lowkey{}. However, this compromises the integrity of the instantiation chain as the 0 potency of the \textit{TextMarker} prohibits the instantiation of \textit{marker\_0}. \lowkey{} resolves this problem by prioritizing the integrity of the instantiation chain and marking \textit{marker\_0} as deleted. While the physical integrity of the model is ensured, linguistic well-formedness might be violated. For example, the \textit{TextMarker} object does not define a \textit{symbol} attribute, which might violate linguistic well-formedness rules. As explained in \secref{sec:framework-linguistic}, such rules are defined by language engineers and tool builders at the linguistic modeling layer.
    \item Should the Language Designer set the potency of the \textit{Marker} class to 2 again, \textit{marker\_0} instance becomes available to the Domain experts again.
\end{itemize}

\input{tables/comparison}

\subsection{Summary}

In this feasibility evaluation, we have demonstrated the usage of \lowkey{} in advanced modeling scenarios.
Concerning the requirements formulated in \secref{sec:case}, we have demonstrated that replicas converge (R1) and that, in a timely fashion (R2). Both causality (R3) and user intention is preserved (R4).
The language engineer's point of view (\secref{sec:elaboration-metamodel}) has shown the development of domain-specific modeling languages with multi-level modeling capabilities (R5).
\lowkey{} enabled reusing the metamodel of the case (\figref{fig:mindmapmm}) that was previously defined in an external tool (R6).
The domain expert's point of view (\secref{sec:elaboration-collaboration}) has demonstrated how real-time collaboration is achieved by CRDT, irrespectively of the level of linguistic abstraction (R5). In addition, we have highlighted how the combination of CRDT and the Phyical metamodel allows for seamless collaboration in the presence of linguistic inconsistencies and non-conformance (R7).

\paragraph{Limitations and Threats to validity.} Our evaluation, being a feasibility study, is limited in some aspects. First, in \secref{sec:elaboration-metamodel}, we used a simple editor developed for our purposes. This setup does not allow for assessing the ease of integration with more complex legacy editors. The main problems that might emerge while integrating with an external editor are primarily associated with architectural choices and the lack of technological convergence. These limitations pose threat to the external validity of our feasibility evaluation. However, these problems are largely mitigated by the relatively loose restrictions the 0MQ-based architecture of \lowkey{} imposes.
  
Second, in \secref{sec:elaboration-metamodel}, we used only three collaborators, while this number might be larger in realistic settings. User intention~\cite{sun1998achieving} becomes harder to ensure as the number of collaborators increases. The lower number of collaborators in our evaluation might threaten conclusion validity. However, these threats are mainly restricted to usability edge cases and do not affect the structural correctness of collaboratively developed models.

Finally, the relatively short example scenarios do not allow observing scalability issues of CRDT caused by the continuously growing number of accumulated messages. However, scalability concerns have not been the focus of this evaluation and we consider the topic future work.

%% file: src/MindMap.py.tex
\begin{lstlisting}[language=Python, caption={Excerpt of the generated model API.}, label={lst:api}]
class MindMap(Clabject):
  # Attribute: title, Type: String,
  # Multiplicity: 1
  def getTitle(self)
  def setTitle(self, title)
  
  # Reference: topic, Type: CentralTopic
  # MultiplicityFrom: 1..1, MultiplicityTo: 1..1
  # IsComposition: True
  def getTopic(self)
  def setTopic(self, topic: CentralTopic)
  def removeTopic(self)
  
  # Reference: markers, Type: Marker
  # MultiplicityFrom: 0..1, MultiplicityTo: 0..*
  # IsComposition: True
  def getMarkers(self)
  def addMarker(self, marker)
  def removeMarker(self, marker)
\end{lstlisting}

%% file: tables/comparison.tex
\begin{table*}[htb!]
\caption{Comparison of related modeling frameworks}
\label{tab:comparison}
\centering
\small
\renewcommand{\arraystretch}{1}

\begin{tabular}{@{}lcccccccc@{}}
\toprule
\multicolumn{1}{c}{\textbf{Framework}} & 
\rot{\textbf{Metamodeling}} &
\rot{\textbf{\begin{tabular}[c]{@{}c@{}}Multi-level\\ modeling\end{tabular}}} &
\rot{\textbf{\begin{tabular}[c]{@{}c@{}}Real-time\\ collaboration\end{tabular}}} &
\rot{\textbf{\begin{tabular}[c]{@{}c@{}}Consistency\\ model\end{tabular}}} &
\rot{\textbf{\begin{tabular}[c]{@{}c@{}}Conflict mgmt/\\ Resolution\end{tabular}}}\\

\midrule

AToMPM \citep{syriani2013atompm} & \harveyBallFull{} & \harveyBallHalf{} & \harveyBallFull{} & Strong & Manual resolution \\
Modelverse \citep{van2017modelverse} & \harveyBallFull{} & \harveyBallFull{} & \harveyBallNone{} & N/A & N/A \\
MetaDepth \citep{delara2010deep} & \harveyBallFull{} & \harveyBallFull{} & \harveyBallNone{} & N/A & N/A \\
Melanee \citep{atkinson2016flexible} & \harveyBallFull{} & \harveyBallFull{} & \harveyBallNone{} & N/A & N/A \\
WebGME \citep{maroti2014next} & \harveyBallFull{} & \harveyBallNone{} & \harveyBallFull{} & Eventual & Manual resolution \\
SpacEclipse \citep{gallardo2012model} & \harveyBallFull{} & \harveyBallNone{} & \harveyBallFull{} & Strong & Manual resolution \\
FlexiSketch \citep{wuest2012flexisketch} & \harveyBallFull{} & \harveyBallNone{} & \harveyBallFull{} & Strong & Prevention \\
SyncMeta \citep{derntl2015near} & \harveyBallHalf{} & \harveyBallNone{} & \harveyBallFull{} & SEC & Prevention (CRDT) \\
MetaEdit+ \citep{kelly2017collaborative} & \harveyBallFull{} & \harveyBallNone{} & \harveyBallNone{} & Strong & Pessimistic locking \\
TGRL \citep{saini2021towards} & \harveyBallNone{} & \harveyBallNone{} & \harveyBallFull{} & SEC & Prevention (CRDT) \\
MONDO \citep{debreceni2017mondo} & \harveyBallNone{} & \harveyBallNone{} & \harveyBallNone{} & Strong & Pessimistic locking \\

\midrule

\lowkey{} & \harveyBallFull{} & \harveyBallFull{} & \harveyBallFull{} & SEC & Prevention (CRDT)

\\ \bottomrule
\end{tabular}

\end{table*}

%% file: sections/discussion.tex
\section{Discussion}\label{sec:discussion}

In this section, we assess how \lowkey{} compares to other modeling and collaborative frameworks. Then, we reflect on various aspects of the approach.

\subsection{Comparison}\label{sec:discussion-comparison}

We compare \lowkey{} with frameworks that are closest in their aim and feature set, shown in~\tabref{tab:comparison}. These are typically either modeling tools with multi-level modeling capabilities or tools with real-time collaborative features. Our objective is to assess how \lowkey{} compares in terms of the key functionality of (i) metamodeling, (ii) multi-level modeling, and (iii) real-time collaboration. We find that some tools partially overlap with \lowkey{} in terms of functionality; however, the combination of the three key features is unique to \lowkey{}.

\textbf{Metamodeling}, i.e., the ability to construct new metamodels and modeling languages, is supported by the majority of the sampled tools.
MetaEdit+ \citep{kelly2017collaborative} is a widely adopted metamodeling framework. Locking at the class level (not attributes) is the primary collaborative mechanism, but there is no support for real-time collaboration. MetaEdit+ represents a class of modeling tools that gained substantial industrial adoption, and could benefit from a real-time collaborative framework such as \lowkey{}.
The two exceptions are TGRL \citep{saini2021towards} and the MONDO framework \citep{debreceni2017mondo}. TGRL is a tool for requirements modeling in a collaborative way. MONDO provides collaborative mechanisms for domain-specific modeling.

\textbf{Multi-level modeling}, i.e., the ability to define metamodels at an arbitrary number of meta-levels, is supported by the mechanisms of deep characterization \citep{atkinson2008reducing} and deep instantiation \citep{atkinson2002rearchitecting}.
AToMPM \citep{syriani2013atompm} is a web-based multi-view modeling tool that allows for defining metamodels at arbitrary number of levels and instantiating them by bootstrapping mechanisms. However, deep characterization and deep instantiation are not supported.  The Modelverse \citep{van2017modelverse} is a modelware back-end for storing and simulating models. It achieves multi-level modeling by using a physical metamodel similar to \lowkey{} \citep{van2014multi}. The same approach has been used by deep metamodeling frameworks MetaDepth \citep{delara2010deep} and Melanee \citep{atkinson2016flexible}.
Similar to \lowkey{}, these tools use graphs at the meta-circular level, \ie the topmost linguistic meta-level.

\textbf{Real-time collaboration} is becoming increasingly adopted in model editors \citep{david2021collaborative}. Tools such as WebGME \citep{maroti2014next}, SpacEclipse \citep{gallardo2012model}, FlexiSketch \citep{wuest2012flexisketch}, and to some extent SyncMeta \citep{derntl2015near} support metamodeling by shallow instantiation, augmented with real-time collaboration capabilities. As a consequence of shallow instantiation, these tools are restricted to a three-level meta-hierarchy, such as OMG's MOF
or EMF \citep{steinberg2008emf}.
A variety of \textbf{consistency models} are employed in these tools to support collaboration. Strong consistency, employed in AToMPM, SpacEclipse, and FlexiSketch, ensures that all participating nodes hold the exact same state of the model at all times. However, due to its underlying mechanisms, it significantly hinders the scalability and user experience of collaborative modeling tools \citep{lamport1978time}. In AToMPM, real-time collaboration is supported in two ways: at the levels of the abstract and concrete syntax. In both cases, changes in the abstract syntax are shared; in the latter case, changes in the representation are shared as well. 
Collaboration in SpacEclipse and AToMPM requires manual \textbf{conflict resolution}, while FlexiSketch uses preventive conflict management techniques.
WebGME relies on eventual consistency, that provides the weaker guarantee that changes will be eventually observed across each node \citep{vogels2009eventually}. However, conflict resolution is not automated. Novel types of real-time collaborative tools, such as TGRL and SyncMeta employ strong eventual consistency (SEC) that combines the benefits of strong and eventual models \citep{shapiro2011comprehensive} and avoids conflicts altogether. TGRL and SyncMeta implement real-time collaboration using the Teletype and Yjs CRDT frameworks, respectively.

\noindent\textbf{Summary.} As shown in \tabref{tab:comparison}, \lowkey{} provides a unique combination of features for real-time collaborative multi-level modeling. 
Typically, modeling frameworks either provide multi-level modeling capabilities without support for real-time collaboration (e.g., Modelverse, MetaDepth); or provide real-time collaboration capabilities without support for multi-level modeling (e.g., WebGME, FlexiSketch). Closest to our work is AToMPM, which provides limited facilities for multi-level modeling, and supports real-time collaboration by conservative consistency model and without automated conflict resolution.

\subsection{Physical metamodel}\label{sec:discussion-physicalmm}

One of the main benefits of the physical metamodel is the uniform representation of objects and models, irrespective of the linguistic meta-level they are situated at. This mechanism allows for the co-existence of models with different syntaxes and semantics. As a consequence, the collaboration between different modeling tools becomes a more manageable endeavor.

We have chosen graphs as the meta-circular level. That is, the physical metamodel corresponds to graphs, and all linguistic models correspond to graphs as well. As discussed in \secref{sec:discussion-comparison}, graphs have been shown to be an appropriate and versatile choice for such purposes.
We have implemented a directed multigraph formalism, \ie edges have an unambiguous source and target vertex, and multiple edges are allowed between vertices. Directed edges enable navigability of associations in linguistic models, and the multigraph nature enables defining multiple different associations between the same pair of classes. Additionally, the physical metamodel supports disconnected graphs. We found this property useful in enabling the temporal tolerance of linguistic inconsistencies.
Additional graph properties can be implemented and enforced, depending on the use-cases to be supported by the framework.

Some scenarios might necessitate extending the physical metamodel with custom-built types. The feasibility of such techniques has been discussed in previous work~\cite{david2022extensible}.

\subsection{Temporal tolerance of linguistic inconsistencies}\label{sec:discussion-tolerance}

The separated physical metamodel allows for collaboration in the presence of linguistic non-conformance (vertical inconsistencies), enabling advanced collaboration scenarios. For example, ensuring the consistency of a model during the collaboration of stakeholders with different expertise might require tolerating linguistic non-conformance. This has been demonstrated in Scenario 2 of \figref{fig:elaboration-process}. Here, the expertise of Client A is the instantiation of objects and the expertise of Client B is organizing dangling objects into models. Already in this simple example, we were able to introduce a non-conformance between the instance model and the metamodel when Client A did not link the newly created instance to the root object immediately. In practical applications, this issue is vastly exacerbated, as the number of stakeholders, domains, formalisms, and tools increases. The CRDT persisting the physical metamodel ensure horizontal consistency \citep{vanherpen2016ontological}, i.e., stakeholders have a consistent view of the system, even if these views are linguistically incorrect. This enables a smooth collaboration in the presence of linguistic non-conformance.

Tool builders can use \lowkey{} for building safer applications with improved inconsistency management mechanisms. As demonstrated, temporal inconsistencies between modeling operations are naturally occurring phenomena in modeling scenarios. The separation of physical and linguistic concerns enables well-formedness and consistency rules to be captured at the most appropriate levels of abstraction. For example, one might constrain their models at the physical level by enforcing conformance to graphs and at the linguistic level to class diagrams. Additionally, the domain-specific API generation capabilities of \lowkey{}—as demonstrated in \secref{sec:elaboration-metamodel}—further improve the safety of modeling tools by allowing for a rich intermediate representation.
Finally, tool integrators might want to consider \lowkey{} as the mediating layer in their tool chains. On the one hand, \lowkey{} is agnostic of the internal representations of the tools to be integrated, allowing to accommodate, in principle, any modeling formalism. On the other hand, the multi-level metamodeling architecture allows for ensuring consistency rules at arbitrary levels of abstraction. The architecture of \lowkey{} makes it especially suitable for developing complex applications and tool chains. Tools to be integrated simply need to establish communication with the \lowkey{} Server through the 0MQ messaging broker used in our architecture. The complexity of this step is substantially mitigated by 0MQ's support for a wide array of programming languages.

\subsection{Support for complex modeling operations}

Throughout this paper, we have assumed atomic change operations (e.g., CRUD operations in the running example). In practical modeling scenarios, more complex domain-specific operations are often required to be supported, e.g., in the automation of large refactoring on the model(s). In the running example, a \textit{promote} operation could be defined to move a \textit{subtopic} with all its children directly under the \textit{central topic}, converting the \textit{subtopic} into a \textit{maintopic} automatically.
The need for such strategies in multi-level modeling has been thoroughly discussed by \citet{delara2018refactoring}. \lowkey{} provides the necessary mechanisms for implementing complex operations in real-time collaborative multi-level modeling settings. While complex modeling operations are defined at the level of the editor, \lowkey{} allows for treating these operations as first-class citizens at the CRDT API level, improving the safety of the software application.

Depending on (i) the synchronicity of message transmission between collaborators, and (ii) the timestamping strategy, there are at least three distinct ways to treat complex operations in a CRDT-based collaborative setting.

First, atomic operations can be aggregated at the client side and transmitted all-at-once as soon as the complex operation is composed, with each atomic operation receiving the same timestamp. This strategy is a natural choice in modeling editors in which atomic operations of a complex functionality are conceptually executed at the same time, with the complex operation being the unit of change. In other words, complex operations are treated as \textit{transactions}. The benefit of this strategy is that the aggregation logic of atomic operations reflects the intentions of the originator: it is the originator who decides which operations to compose into a complex operation.

Second, similarly to the previous alternative, atomic operations can be aggregated at the client side and transmitted all-at-once, however, with each atomic operation receiving its own timestamp. Conceptually, the transmission of atomic operations is delayed until a complex operation is built up at the client side. This allows for beneficial mechanisms at the originator's side, such as local undo-redo and rollback. At the receiving end, atomic operations—--each with its own timestamp--—have to be merged into the local replica. However, the notion of aggregation is lost, unless the clients have previously agreed on the same aggregation semantics.

Finally, atomic operations can be transmitted immediately and aggregated at the receiving clients' side. Clearly, this is the opposite of the first alternative where the originator controls the aggregation. Here, it is the receiving clients who decide what to aggregate and how. This strategy is often used in stream processing and complex event processing~\cite{david2018foundations}, where clients might be interested in their own, domain-specific complex phenomena. For example, a safety engineer who is part of a collaboration might want to execute safety checks after specific complex sequences of changes, but that complex sequence might not interest others.

\subsection{Opportunities in multi-view modeling}

We anticipate multi-view modeling (MVM) \citep{reineke2014basic} being one of the main application areas of our approach. Views are projections of one or multiple underlying models \citep{corley2016cloud}, presenting stakeholders with only the essential information they require for their work. Views typically pertain to domains and expertise (e.g., the mechanical and the electrical views in the design of a mechatronic system), but they can pertain to specific use cases (e.g., electro-mechanic view) or expertise (e.g., chassis design).
The architecture and services of \lowkey{} align well with the requirements of MVM. It is able to accommodate multiple different metamodels, their instances, and their views in a uniform fashion; thus allowing for change propagation between linguistically and semantically different views. Model- and screen sharing \citep{van2018unifying} are straightforward to implement, as tool builders can outsource the data layer of their tools to \lowkey{}.

Other approaches relying on an ensemble of multiple models---such as multi-paradigm modeling (MPM) \citep{mosterman2004computer}, multi-modeling \citep{boronat2008what} and blended modeling \citep{david2022blended}---can benefit from this approach as well. MPM advocates modeling every aspect of the system at the most appropriate level(s) of abstraction using the most appropriate formalism(s). The physical metamodel provides a basis for synchronization among stakeholders, while different formalisms can be implemented at the linguistic level. Similar techniques have been employed in the MPM tool Modelverse \citep{van2017modelverse}.

\subsection{Accommodating mainstream modeling frameworks}

As demonstrated by \citet{atkinson2002rearchitecting}, multi-level modeling frameworks can emulate traditional modeling frameworks using the notion of potency. Potency is a constraint that specifies how many times a class can be instantiated transitively. As shown in \figref{fig:physical-metamodel-crdt}, the physical metamodel of \lowkey{} defines a potency to its \textit{Clabject} element.
By that, \lowkey{} subsumes traditional modeling frameworks that operate on the mechanism of shallow instantiation.
We see an opportunity in developing libraries for \lowkey{} implementing the meta-facilities of these traditional frameworks. The process of adopting \lowkey{} in already existing modeling tools relying on such frameworks, e.g., MOF or EMF, can be vastly improved and automated.

\subsection{Limitations}

\textbf{Performance.} The performance of CRDT is subject to the number of objects present in the specific application \citep{sun2006operation}. \lowkey{} CRDT implement a soft delete mechanism, i.e., objects are never removed from the specific CRDT, but rather, ``marked'' as removed. The \lwwset{}, for example, contains the removed elements in its remove-set (see \secref{sec:framework-crdt}). As a consequence, the performance of the CRDT layer will gradually decrease as the number of objects increases. Our preliminary measurements show linear degradation. Garbage collection mechanisms have been suggested for managing such limitations of CRDT-based applications \citep{bauwens2019memory}.

\noindent{}\textbf{Intention preservation} \citep{sun1998achieving} plays a key role in achieving an intuitive human-computer interaction and a smooth user experience in collaborative settings. While the LWW paradigm usually preserves the user's intention, some corner cases might result in model changes that are less intuitive. The user makes decisions on changing the model based on the model's materialized view in the modeling tool or browser. However, there might be change updates from other collaborators on their way that might arrive after the user carried out his changes. These updates retroactively change the model in a way that affects the user's reasoning. Since there is no way to account for such messages, the user's intention in such corner cases cannot be guaranteed. While manual conflict resolution would solve this issue, it would also render real-time collaboration infeasible.

\noindent{}\textbf{Timestamping mechanism.} The current prototype implementation of \lowkey{} uses the \texttt{time.time_ns()} Python function for timestamping changes. This function returns time as an integer number of nanoseconds. However, this approach is prone to clock drift, which could render CRDT inconsistent. Our working assumption is that nanosecond-level drift does not affect the dynamics of collaborative modeling between humans, as the interactions in such settings are at the level of seconds. Nonetheless, \lowkey{} can be extended by additional mechanisms to regularly synchronize clocks using the Network Time Protocol (NTP)\footnote{\url{http://www.ntp.org/}} and other mechanisms described in \secref{sec:timestamps}.

%% file: sections/conclusions.tex
\section{Conclusion}\label{sec:conclusions}
In this paper, we have presented a real-time collaborative framework for a wide range of advanced modeling scenarios, supported by techniques of multi-level modeling. Our framework provides a unique combination of modeling capabilities and real-time collaboration. It is built on a custom implementation of CRDT, geared towards graph models, providing promising real-time capabilities and scalability. We have defined a mapping of physical metamodels onto CRDT and demonstrated the approach through an illustrative case.

We have identified multiple lines of future work. We are planning to develop state-of-the-art collaborative multi-view modeling mechanisms based on the framework. To enable easier integration with legacy models, we will provide profiles for UML and EMF models, effectively reproducing the respective meta-levels of the OMG superstructure and Ecore. We plan to build a family of modeling editors integrated with existing modeling frameworks to augment them with real-time collaborative capabilities.
In more technical terms, we are looking to improve the performance of the framework by implementing advanced garbage collection mechanisms and a network stack with less overhead. We will develop a benchmark for collaborative MDSE frameworks to evaluate their scalability (with respect to connected clients and messages exchanged), performance (response time of a local operation being propagated to each remote client), and usability in typical modeling scenarios.

%% file: sections/acknowledgments.tex
\section*{Acknowledgments}

The authors would like to express their gratitude for the insightful comments of the reviewers that helped improve the initial manuscript.